\documentclass{iopartbfsubs}
\usepackage{color,times}

\newcount\Hour
\newcount\HM
\newcount\Minute
\Hour=\time\divide\Hour by 60
\Minute=\time
\HM=\Hour\multiply\HM by 60
\advance\Minute by -\HM
\ifnum\Minute=0\def\Mt{00}\fi%
\ifnum\Minute>0\def\Mt{0\the\Minute}\fi%
\ifnum\Minute>9\def\Mt{\the\Minute}\fi%
\def\Now{\the\Hour .\Mt}

\renewcommand{\tr}[2]{\mathop{{\rm Tr}_{#1}\left\{#2\right\}}}
\newcommand{\Label}[1]{\label{#1}}
\newcommand{\cwg}[1]{}

\def\DRAFT{%
\renewcommand{\cwg}[1]{{\color{blue}##1}}%
\renewcommand{\Label}[1]{\label{##1}
{\hbox to 0cm{\textcolor{magenta}{\hss\em ##1\quad}}}}}

\newcommand{\QSDE}{quantum stochastic differential equation}
\newcommand{\QQSDE}{Quantum stochastic differential equation}
\newcommand{\<}{{\langle}}
\renewcommand{\>}{{\rangle}}
\renewcommand{\d}{{\delta}}
\newcommand{\p}{\pi}
\newcommand{\w}{\omega}
\renewcommand{\t}{\tau}
\renewcommand{\k}{\kappa}

\newcommand{\D}{\Delta}
\newcommand{\Ito}{\mbox{\bf{(I)\,}}}
\newcommand{\Strato}{\mbox{\bf{(S)\,}}}
\newcommand{\half}{{1\over 2}}

\begin{document}
\title[Input and output in damped quantum systems III: Fermionic..]{Input and 
output in damped quantum systems III:
 Formulation of damped systems driven by Fermion 
fields}

\author{C. W. Gardiner}
\address{School of Chemical and Physical Sciences, Victoria 
University of Wellington, New Zealand
\ 
\\ {\em\today\ at \Now}}

\begin{abstract}

A comprehensive input-output theory is developed for Fermionic input 
fields. Quantum stochastic differential equations are developed in 
both the Ito and Stratonovich forms.
The major technical issue  is the development of a formalism 
which takes account of anticommutation relations between the 
Fermionic driving 
field and those system operators which can change the number of 
Fermions within the system.
\end{abstract}

\section{Introduction}
\Label{sec:intro}
The introduction of Input-Output formulations in the 
1980s~\cite{Collett1984a,Gardiner1985a,Gardiner1987a} was a response 
to the necessity for a theory of quantum damping which could deal with 
travelling wave situations.  A formulation of a photodetector theory 
was suggested using input-output methods \cite{Gardiner1989a,QN}, in 
which the detection of a photon is envisaged as the conversion of an 
input light field into an output electron field.  However, to do this 
requires a theory of inputs and outputs with Fermion fields, and an 
elementary theory was developed.
Although basically satisfactory, this formulation was not complete, 
and in particular, could not be used to derive a master equation.  
The problem that arises is quite simple: the equations of motion for 
system generators are different, depending on whether a system 
operator is viewed as {\em commuting\/ } or {\em anticommuting\/} 
with the fermionic heat bath operators.

In this paper this technical problem is overcome.  It is shown how we 
may define ``restricted'' system operators so as to 
commute with all bath operators, and that these internal system 
operators all obey a quantum Langevin equation of the same form.

However, the equations of motion in the original operators are, for 
the two level atom, linear and thus exactly soluble.  Thus, we have 
a description of a two level system interacting with a Fermi Bath 
which is essentially the same as that of a harmonic oscillator 
interacting with a Bosonic heat bath.  The two level atom behaves 
like a Fermion coming to equilibrium with all the other Fermions.

It is possible to develop Fermionic quantum stochastic integration, 
and the corresponding Ito and Stratonovich formulations of \QSDE s, 
and finally, from these, to derive the Master equation in the 
expected form.

To the best of the author's knowledge there have only been two other 
papers dealing with input-output theory of Fermions, that of Sun and 
Milburn~\cite{Sun1999a} and that of Search~\etal~\cite{Search2002a}.  
Neither of these attempts a comprehensive input-output formalism; the 
first concentrates essentially on counting statistics, while the focus 
of the second paper is on the theory of Fermions inside and outside of 
a linear cavity, and does not address the quantum stochastic issues 
which are the principal topic of this paper.
\section{Beams of Fermions}
\Label{sec:BF}
\subsection{Input and Output Fields}
We want to consider here non-relativistic Fermi fields, in which we 
shall for simplicity and clarity make no mention of spin (though 
this is always present in Fermion, it plays no essential r\^ole in 
their non-relativistic description).  For simplicity, we shall also 
consider propagation in only one dimension.  A Fermi field can then 
be written:
\begin{eqnarray}
\Label{2.1}
D(t,x) &=& \sqrt{{1\over2\pi  }} \int^\infty_{-\infty} dk\, 
\rme^{\rmi (kx-\omega t)} d(k)
\end{eqnarray}
where $ d(k)$ is the destruction operator, with anticommutation 
relations
\begin{eqnarray}
[d (k),d^\dagger(k')]_+  &=&\delta(k-k')
\nonumber \\
\Label{2.2}
[d (k),d (k')]_+   	&=& [d^\dagger(k),d^\dagger(k')]_+ = 0 
\end{eqnarray}
leading to the equal time anticommutation relation,
\begin{eqnarray}
\Label{2.3}
[D(t,x), D^\dagger(t,x')]_+ &=& \delta(x-x').
\end{eqnarray}
The dispersion relation
\begin{eqnarray}
\Label{2.4}
\omega = \hbar k^2/2m
\end{eqnarray}
follows from non-relativistic mechanics, and ensures $ D(t,x)$ obeys 
the time dependent Schr\"odinger equation
\begin{eqnarray}
\Label{2.5}
\rmi\hbar{\partial D(t,x)\over\partial t } = - {\hbar^2\over2m  }
{\partial ^2 D(t,x)\over\partial x^2  }
\end{eqnarray}
\cwg{More explanation needed}\\
In the situation we wish to consider an input field radiating into 
a system, which itself radiates an output field in the opposite 
direction.  In that case, it is more appropriate to consider the 
field as being defined on the half-line $ 0 \leq x < \infty$, for 
which an appropriate expansion is
\begin{eqnarray}
\Label{2.6}
D(t,x) = \sqrt{{2\over\pi  }} \int^\infty_0 dk \cos (kx) \rme^{\rmi  
\omega t} d(k) 	
\end{eqnarray}
which can also be written
\begin{eqnarray}\Label{2.7}
D(t,x) = \sqrt{{1\over 2\pi }}
 \int^\infty_0 dk \bigl\{\rme^{\rmi (kx-\omega t)} d_{\rm in}(k) 
+ \rme^{-\rmi (kx +\omega t)} d_{\rm out}(k)\bigr\}
\end{eqnarray}
where
\begin{eqnarray}
\Label{2.8}
d_{\rm in}(k) = d_{\rm out} (k) = d(k) .
\end{eqnarray}
We can also define
\begin{eqnarray}
\Label{2.90}
D_{\rm in}(t,x)	   	&=&	\sqrt{{1\over 2\pi }} \int^\infty _0
dk\,\rme^{-\rmi (kx-\omega t)} d_{\rm in} (k)\nonumber 	\\
\Label{2.9}
D_{\rm out} (t,x)  	 &=& \sqrt{{1\over 2\pi }} \int^\infty_0 
dk\,\rme^{-\rmi (kx+\omega t)} d_{\rm out} (k)
\end{eqnarray}
and the boundary condition
\begin{eqnarray}
\Label{2.10}
D_{\rm in}(t,0) - D_{\rm out} (t,0) = 0.
\end{eqnarray}
follows from (\ref{2.8}).
\subsection{Dispersion}
Matter waves are dispersive; in fact the group and phase velocities 
differ by a factor of 2 for all frequencies.  This means that the 
simple propagation of statistics as in light beams is not valid 
here---there is no solution analogous to the solution of the one 
dimensional wave equation which can be written in the form
\begin{eqnarray}
\Label{2.11a}
A(t,x) = A_{\rm in}(t + x/v).
\end{eqnarray}
We will therefore limit our considerations to very narrow bandwidth 
situations.  In order to judge what sort of bandwidth can be 
considered ``narrow'', let us consider the correlation functions.

\subsection{Correlation Functions of Propagating Fermion Beams}
We want to consider correlation functions
\begin{eqnarray}
\Label{2.11b}
G^{(1)} (x_1,t_1; x_2,t_2) = \< D^\dagger(t_1,x_1) 
D(t_2,x_2)\>
\end{eqnarray}
and
\begin{eqnarray}\fl\Label{2.12}
G^{(2)} (x_1,t_1;x_2,t_2; x_3,t_3; x_4,t_4)
&=& \< D^\dagger(x_1,t_1)D^\dagger(x_2,t_2) D(x_3,t_3)
D(x_4,t_4)\> 
\end{eqnarray}
These are analogous to the similarly defined correlation functions 
for optical fields.

Let us now consider a stationary narrow bandwidth field, such that
\begin{eqnarray}
\Label{2.13}
\< a^\dagger(k) a(k')\> = \bar{N}(k) \delta(k-k')
\end{eqnarray}
where
\begin{eqnarray}
\Label{2.14}
\bar{N}(k) = 0  \qquad 	\hbox{unless} \qquad k \approx k_0  .
\end{eqnarray}
Now defining
\begin{eqnarray}
\Label{2.15}
N(\omega)d\omega = \bar{N}(k) dk
\end{eqnarray}
then
\begin{eqnarray}
\Label{2.16}
\< D^\dagger(t,x) D(t',x')\> = {1\over2 \p  } \int d\omega\,N(\w)
\rme^{\rmi \omega[(t-t') - k(x-x')/\omega]}
\end{eqnarray}
If $ \omega_0, k_0$ are the central frequency and wavenumber of the 
range, and the central velocity of propagation is
\begin{eqnarray}
\Label{2.17}
v = \omega_0/k_0
\end{eqnarray}
then, provided that the range of frequencies, $ \delta\omega$, 
satisfies
\begin{eqnarray}
\Label{2.18}
(x-x')\,\delta\omega/v \ll 1
\end{eqnarray}
we can first write
\begin{eqnarray}
\Label{2.19}
g(\t) = \rme^{-\rmi \omega_0\t} \< D^\dagger(\t,x) D(0,x)\>
\end{eqnarray}
and then we can derive
\begin{eqnarray}\fl\Label{2.20}
G^{(1)}(   	x,t; x',t')	
   	&=& \rme^{\rmi \omega_0[t-t'-(x-x')/v]} 
\left(1 + {x-x'\over2v  }{\partial\over\partial t  }\right)
g\left(t-t'-{x-x'\over v  }\right)
\end{eqnarray}
This equation gives the correction due to dispersion of the 
``propagation approximation'' to evaluating the correlation function 
of a fermion beam at two separated points.  The condition of validity 
for such an approximation is (\ref{2.18}), which can be 
interpreted to mean that the size of individual wavepackets, which 
is order of magnitude $ v/\delta\omega$, must be very much larger 
than the distance $ |x-x'|$ between the points considered.

The correction in (\ref{2.20}) is relevant to the measurement of 
time correlation functions by delayed coincidence measurements, when 
the delay is induced by allowing one beam to propagate further than 
the other.

\subsection{Thermal correlation functions for Fermionic beams}
In the optical case a thermal light beam is considered to be 
Gaussian, and the factorizable property of Gaussian moments leads to 
a relationship between 1st and 2nd order correlation functions which 
is the characteristic of the ``bunched'' nature of thermal light. It 
is difficult to define what might be considered to be a ``Gaussian'' 
Fermion state, but a {\em thermal\/} Fermion state can be defined.

In this case of a thermal state we have
\begin{eqnarray}
\fl\Label{2.21}
\< d ^\dagger(k) d(k')\> &=& \delta(k-k') \bar{N}(k)
\\ \fl \Label{2.22}
\< d^\dagger(k)	d^\dagger(k')d(k'')d(k''')\>
&=&
\left[\d(k-k'')\d(k'-k''')-\d(k-k''')\d(k'-k'')\right]
\bar{N}(k)\bar{N}(k').
\nonumber\\
\end{eqnarray}
Here, if
\begin{eqnarray}
\Label{2.23}
\bar{N}(\omega)\,d\omega = \bar{N}(k)dk,
\end{eqnarray}
then
\begin{eqnarray}
\Label{2.24}
\bar{N}(\omega) = 1/\left[\exp(\hbar\omega/kT) + 1\right].
\end{eqnarray}
The antisymmetric requirement (\ref{2.22}) is in fact the natural 
analogy of the Gaussian factorization property of photon 
beams (there would be a $ +$ sign on the RHS of (\ref{2.22}) in the case 
of photons instead of a $ -$ sign.)  It leads to the relationship
\begin{eqnarray}\fl\Label{2.25}
G^{(2)}(   	x_1,t_1; x_2,t_2;x_3,t_3;x_4,t_4)  
   	   	   	&=&G^{(1)}(x_1,t_1;x_3,t_3)G^{(1)}(x_2,t_2;x_4,t_4)	   	   
\nonumber \\\fl
   	&&- G^{(1)}(x_1,t_1;x_4,t_4) G^{(1)}(x_2,t_2;x_3,t_3).
\end{eqnarray}
The corresponding formula for a Gaussian Boson beam differs only by 
having a positive sign rather than a negative sign on the right hand 
side.

In the case of stationary statistics, evaluated with time difference 
$ \t$ at $ x=0$, we have for
\begin{eqnarray}
\Label{2.26}
g^{(2)}(\t) \equiv G^{(2)}(0,t; 0,t+\t; 0,t+\t;0,0)
\end{eqnarray}
\begin{eqnarray}
\Label{2.27}
g^{(1)}(\t) = G^{(1)} (0,t; 0,t+\t)
\end{eqnarray}
that
\begin{eqnarray}
\Label{2.28}
g^{(2)}(\t)	   	&=& g^{(1)}(\t) g^{(1)}(-\t )-[g^{(1)}(0)]^2
  	\\
\Label{2.29}
   	   	   	   	&=& |g^{(1)}(\t)|^2 - |g^{(1)}(0)|^2
\end{eqnarray}
Clearly $ g^{(2)}(0) = 0$, corresponding to perfect antibunching, as 
expected from a Fermion beam.


\section{Interaction of a System with a Fermionic heat bath}
\Label{sec:interaction}
In order to fix our ideas, let us consider the ionization of an atom 
under the influence of an impingent electron beam.  The Hamiltonian 
is
\begin{eqnarray}
\Label{3.1}
H=H_{\rm sys} + H_{\rm Int} + H_B.
\end{eqnarray}
Here, $ H_{\rm sys}$ is the free atom Hamiltonian, whose precise 
form will be left open.  The bath Hamiltonian, corresponding to a 
field on a {\em half line\/}, $ 0 < x < \infty$, can be written
\begin{eqnarray}
\Label{3.2}
H_B = \int^\infty_0 dk\,\hbar\omega(k)\,d^\dagger(k)d(k)
\end{eqnarray}
where in this case, for an electron of mass $ m$,
\begin{eqnarray}
\Label{3.3}
\omega(k) = \hbar k^2/2m.
\end{eqnarray}
The Fermion operators $ d(k)$ are as in the previous section.  
Finally, the interaction is conceived as representing the absorption 
or emission of an electron, and is written
\begin{eqnarray}
\Label{3.4}
H_{\rm Int}=\rmi\hbar\int^\infty_0 dk\,\k(k)
\bigl\{d^\dagger(k)\tilde{c} - \tilde{c}^\dagger d(k)\bigr\}
\end{eqnarray}
Here $ \tilde{c},\tilde{c}^\dagger$ are system operators.  The 
action of $ \tilde{c}^\dagger$ on a system state increases the 
number of constituent electrons by 1.
\subsection{Fermionic and Bosonic System Operators}
The commutation relations for the system operators depend on the 
systems being studied.  In the usual case of a Boson bath [1] the 
system and bath operators {\em commute\/} at equal times.  However, 
when dealing with a Fermionic bath the situation is different, 
depending on whether a system operator can be considered as changing 
the number of Fermions which make up the system or not.  The 
separation into ``{\em system\/}'' and ``{\em bath\/}'' tends to obscure the 
fact that the system does have an internal structure, and that (for 
example) an ion and a neutral atom must have different numbers of 
constituent electrons.  An operator such as $ \tilde{c}$, which can 
be regarded as removing an electron from the neutral atom, must 
anticommute with all bath operators, since it must be composed of an 
odd number of creation and destruction operators.  On the other 
hand, there are operators (such as $ H_{\rm sys}$) which do not 
change the number of constituent electrons, and hence commute with 
the bath operators at equal times.  We thus conclude that these are 
two kinds of system operators.
\begin{itemize}
\item[a)] Bosonic---these commute with all bath operators, $ d(k),d^
\dagger(k)$ (at equal times)
\item[b)] Fermionic---these anticommute with all bath operators $ d(k),d^
\dagger(k)$ (at equal times)
\end{itemize}
We will use the notation $ \tilde{a}, \tilde{b}, \tilde{c}$, etc. 
for the system operators to emphasize that such operators {\em may} 
anticommute with the bath operators, and hence are not necessarily 
independent of them.  We will shortly introduce ``restricted'' 
system operators, which are independent of and hence commute with the bath 
operators.
\subsection{Derivation of Quantum Langevin Equations}
Because all Fermion fields of interest are massive, and therefore 
the wave propagation is dispersive it is not quite as easy to derive 
quantum Langevin equations as in the optical case.  Added to this is 
the complication that some of the system operators are Fermionic, 
and others Bosonic.

The operator $ \tilde{c}$ which occurs in the interaction 
Hamiltonian must be Fermionic, since it changes the number of 
constituent electrons in the system by 1.  Using this, we derive the 
equation of motion for $ d(k)$.
\begin{eqnarray}
\Label{3.5}
\dot{d}(k) = - \rmi \omega (k) d(k) - \k\tilde{c}
\end{eqnarray}
which we can integrate to get
\begin{eqnarray}\Label{3.6}
d(k,t)  = \rme^{-\rmi \omega (k)(t-t_0)} d(k,t_0) - {\k}(k)\int^t_{t_0}
\rme^{-\rmi \omega (k)(t-t')} \tilde{c}(t')dt'
\end{eqnarray}
We now define
\begin{eqnarray}
\Label{3.7}
d(t)&\equiv&  {1\over 2 }D(t,0)	   	= \sqrt{{1\over 2\p }}
\int^\infty_0 dk\,d(k,t)   	\\
\Label{3.8}
d_{\rm in}(t)  	   	   	   	   	   	&=& \sqrt{{1\over  2 \p}}
\int^\infty_0dk\,\rme^{-\rmi \omega(k)(t-t_0)}d(k,t_0)
\end{eqnarray}
The quantity $ d(t)$ is a genuine Heisenberg operator for the time $ 
t$, whereas $ d_{\rm in}(t)$ depends only on the initial values $ 
d(k,t_0)$, of the destruction operators.  We now write the equation 
of motion for an arbitrary system operator $ \tilde{a}$,
\begin{eqnarray}
\Label{3.9}
\dot{\tilde {a}}   	&=& -{\rmi\over \hbar }[\tilde{a},H_{\rm sys}]
+ \int^\infty_0 dk\,{\k}(k) \left[\tilde{a},d^\dagger(k)
\tilde{c} - \tilde{c}^\dagger(k)d(k)\right]	\\
\Label{3.10}
&=& -{\rmi\over \hbar }\left[\tilde{a},H_{\rm sys}\right]
\nonumber\\ 
&&+ \int^\infty_0 dk\, {\k}(k) \bigl\{\mp d^\dagger (k)[\tilde{a},
\tilde{c}]_\pm-[\tilde{a},\tilde{c}^\dagger]_\pm d(k)\bigr\}
\end{eqnarray}
where the {\em top\/} signs apply if $ \tilde{a}$ is Fermionic, and 
the bottom signs apply if $ \tilde{a}$ is Bosonic.
\subsubsection{The white noise approximation}
To obtain Langevin equations we must make approximations.  There are 
two principal approximations.  
\begin{itemize}
\item[a)] The interaction is 
weak, and the free motion of $ \tilde{c}(t)$ is proportional to $ 
\rme^{-\rmi \omega_0t}$. 
\item[b)] The frequency $ \omega_0$ is rather large.  
Since the equation (\ref{3.10}) is homogeneous in $ \tilde{a}$, this 
means that the main contribution from the integrals will occur where 
$ \omega(k )\approx \omega_0$. 
\end{itemize}
We can thus write an approximate 
expression for the $ k$ integrals in (\ref{3.10}) by evaluating the 
c-number $ \k(k)$ at $ k_0$; i.e.,
\begin{eqnarray}
\Label{3.11}
\int^\infty_0 dk\,\k(k)d(k) \approx \k(k_0)\int^\infty_0 dk\,d(k) 
\approx
\sqrt{2\p}\,{\k}(k_0)\,d(t).
\end{eqnarray}
provided $ \k(k)$ is a smooth function of $ k$ around $ 
\omega (k) = \omega_0$.  From (\ref{3.6})
\begin{eqnarray}
\Label{3.12}
d(t)   	&=& d_{\rm in}(t) + \sqrt{{1\over 2\p }}\int^\infty_0
dk'\,\k(k') \rme^{-\rmi \omega (k)(t-t')}\tilde{c}(t')\,dt'	\\
\Label{3.13}
   	   	&=& d_{\rm in}(t) + \sqrt{1\over 2\p } \int^t_{t_0}
dt' \int d\omega {dk(\omega)\over d\omega }\k(k(\omega))
\rme^{-\rmi \omega(t-t')} \tilde{c}(t')\nonumber\\ 
\end{eqnarray}
and provided $ \k(k(\omega))$ and $ dk(\omega)/d\omega$ are 
smooth around $ \w=\w_0$, we can again approximate:
\begin{eqnarray}
\Label{3.14}
d(t) &=& d_{\rm in} 
 + \sqrt{2\p}\int^t_{t_0}
\delta (t-t')\tilde{c} (t')
\left[ \k\big(k(\omega)\big) {dk(\omega)\over d\omega }\right]
_{\omega=\omega_0}
\end{eqnarray}
that is
\begin{eqnarray}
\Label{3.15}
d(t) = d_{\rm in}(t) + \sqrt{{\p\over 2 }}\k
\big(k(\omega_0)\big){dk(\omega_0)\over d\omega_0 }\tilde{c}(t)
\end{eqnarray}
Before finally substituting to derive the quantum Langevin equations, 
notice that
\begin{eqnarray}
\Label{3.15a}
\left[d_{\rm in}(t),d^\dagger _{\rm in}(t')\right]_+
   	   	&=& {1\over 2\p }\int^\infty_0 dk\,\rme^{-\rmi \omega(k)(t-t')}
\\
\Label{3.16}
   	   	&=& {1\over 2\pi } \int^\infty_0 d\omega 
{dk(\omega)\over d\omega }  \rme^{-\rmi \omega(t-t')}
\end{eqnarray}
and if this is being used mostly at frequency $ \omega_0 \gg 0$, 
then we can approximate to get
\begin{eqnarray}
\Label{3.17}
\left[d_{\rm in}(t),d^\dagger _{\rm in}(t')\right]_+ =
{dk(\omega_0)\over dk  } \delta (t-t')
 \end{eqnarray}
We  want a noise input with anticommutator normalized to $ 
\delta (t-t')$; we therefore define
\begin{eqnarray}
\Label{3.18}
f_{\rm in}(t) = \left[{dk(\omega_0)\over d\omega_0  }\right]^{-{1 
\over 2}} d_{\rm in}(t) ,
\end{eqnarray}
so that (\ref{3.11}) becomes
\begin{eqnarray}\Label{3.1801}
\int_0^\infty dk\, \kappa(k) d(k) 
&\approx& {\gamma\over 2}\tilde c + \sqrt{\gamma}\,f_{\rm in}(t),
\end{eqnarray}
where
\begin{eqnarray}
\Label{3.20}
\gamma = 2\pi\k\big(k(\omega_0)\big) ^2
   	   	   	   	   	   	   	\left({dk(\omega_0)\over d\omega_0  }\right).
\end{eqnarray}
Using this approximation, the interaction Hamiltonian can also be approximated 
by
\begin{eqnarray}\Label{3.2001}
H_{\rm Int}&\approx& 
\rmi\hbar\left\{
\left({\gamma\over 2}\tilde c + \sqrt{\gamma}\,f^\dagger_{\rm in}(t)\right)
\tilde c
-\tilde c^\dagger\left({\gamma\over 2}\tilde c + \sqrt{\gamma}\,f_{\rm in}(t)
\right)\right\},
\end{eqnarray}
and this is a form which will be useful in the remainder of this paper.

\subsubsection{Fermionic quantum Langevin equations}
Now substitute into (\ref{3.10}) to get
\begin{eqnarray}\fl\Label{3.19}
\dot{\tilde{a}}& =&
 - {\rmi\over \hbar  }\left[\tilde{a},H_{\rm sys}\right]   	   
-\left[\tilde{a},\tilde{c} ^\dagger \right]_\pm 
\left\{{\gamma\over 2 }\tilde{c} + \sqrt{\gamma}f_{\rm in} (t)\right\} 	
\mp \left\{{\gamma\over 2 }
\tilde{c} ^\dagger  + \sqrt\gamma f ^\dagger _{\rm in}(t)\right\}[\tilde{a},
\tilde{c}]_\pm
\end{eqnarray}
Equations (\ref{3.19}) are the Fermionic quantum Langevin equations 
for the full system operators.  At this stage all we know is that 
the $ f_{\rm in}(t)$ are determined by the $ d(k,t_0)$, whose 
statistics are determined from the initial state of the Fermionic 
heat bath, and that
\begin{eqnarray}
\Label{3.21}
\bigl[f_{\rm in}(t),f^\dagger_{\rm in}(t')\bigr]_+ =
\delta(t-t').
\end{eqnarray}
The validity of (\ref{3.19}) and (\ref{3.21}) is restricted to 
situations in which the interaction is rather weak, and in which the 
time dependence of $ \tilde{c}(t)$ in the case of no interaction is 
$ \rme^{-\rmi \omega_0t}$, for some rather large $ \omega_0$.


\subsubsection{``Out'' operators}
As in the case of Bosonic quantum white noise, we can define ``out'' 
operators, by considering solutions of (\ref{3.5}) in terms of a {
\em final\/} condition at a time $ t_1 > t$.  Thus, (\ref{3.6}) 
becomes 
\begin{eqnarray}
\Label{3.2101}
d(k,t) = \rme^{-\rmi\omega (k)(t-t_1)}d(k,t_1) + \int^{t_1}_t
{\k}(k) \rme^{-\rmi\omega (k)(t-t')}\tilde{c}(t')dt'.
\end{eqnarray}
and we define
\begin{eqnarray}
\Label{3.2102}
d_{\rm out} (t) = \sqrt{{1\over 2\p }}\int^\infty_0
dk\,\rme^{-\rmi\omega (k)(t-t_1)} d(k,t_1 ).
\end{eqnarray}
and from (\ref{3.2102}) and (\ref{3.13})
\begin{eqnarray}
\Label{3.2103}
d_{\rm out}(t) &-& d_{\rm in}(t) \nonumber\\ 
&=& \sqrt {1\over 2\p } 
\int^{t_1}_{t_0} dt' \int^\infty_0 dk\,{\k}(k) 
\rme^{\rmi\omega (k)(t-t')}\tilde{c}(t')dt'
\end{eqnarray}
and using the same methods as in (\ref{3.14}) to (\ref{3.15})
\begin{eqnarray}
\Label{3.2104}
d_{\rm out}(t) - d_{\rm in}(t) = \sqrt{2\p} {\k}\big(k(\omega_0)\big)
{dk(\omega _0)\over d\omega _0 }\tilde{c}(t).
\end{eqnarray}
or, in terms of $ f_{\rm in}(t),f_{\rm out}(t)$,
\begin{eqnarray}
\Label{3.2105}
f_{\rm out}(t) - f_{\rm in}(t) = \sqrt{\gamma}\,\tilde{c}(t).
\end{eqnarray}
We can derive ``time reversed'' quantum Langevin equations in terms 
of these ``out'' noises.
\subsubsection{Commutation Relations between System and Input 
Operators}
The operators $ d(t),d^\dagger(t)$, defined by (\ref{3.7}), commute 
(anticommute) with all Bosonic (Fermionic) system operators at the 
same time, since these describe independent degrees of freedom.  
Thus, using (\ref{3.15}) and (\ref{3.18}), we can say that if 
$ \tilde a(t)$ is an arbitrary system operator
\begin{eqnarray}\Label{3.2106}
[\tilde a(t),d(t)]_\pm=	0
&  \quad \Longrightarrow \quad &[\tilde a(t),f_{\rm in}(t)]_\pm
=-{1\over 2 }\sqrt{
\gamma}
[\tilde a(t),\tilde{c}(t)]_{\pm}
\end{eqnarray}
If we rewrite (\ref{3.2105}) as
\begin{eqnarray}
\Label{3.2107}
f_{\rm in}(t) + {1\over 2 }\sqrt{\gamma}\,\tilde{c}(t)
=f_{\rm out}(t) - {1\over 2 }\sqrt{\gamma}\,\tilde{c}(t)
\end{eqnarray}
it is easy to rewrite the quantum Langevin equation (\ref{3.19}) in 
the ``out'' form
\begin{eqnarray}
\dot{\tilde{a}} = -{\rmi\over\hbar  }
\left[\tilde{a},H_{\rm sys}\right] 
   	&-& \left[ \tilde{a},\tilde{c}^\dagger \right]_{\pm}
\bigl\{-{\gamma\over 2 }\tilde{c} + \sqrt{\gamma}f_{\rm out}(t)
\bigr\}	   	\nonumber \\
\Label{3.2108}
   	&\mp&\bigl\{-{\gamma\over 2 }\tilde{c}^\dagger + \sqrt{\gamma}
f^\dagger_{\rm out}	(t) \bigr\} [\tilde{a},\tilde{c}]_\pm.
\end{eqnarray}
\subsection{Restricted System Operators}
Because the bath operators do not commute with those system 
operators $ \tilde{a}$ which are Fermionic, we have different forms 
for the quantum Langevin equation depending on whether or not the 
operator under consideration is Fermionic.

We shall introduce a different set of system operators, called {\em 
restricted\/} system operators, which do not have this problem,
that is, the equations of motion take the same form for all restricted system 
operators.

To  do this we introduce the operator $ I$, in the bath space, which 
anticommutes with all bath operators.
\begin{eqnarray}
\Label{3.22}
[I,d(k)]_+ = [I,d^\dagger(k)]_+ = 0
\end{eqnarray}
This operator is easy to construct explicitly.  If $ |A\> $ is any 
bath state with a definite number $ n_A$ of bath Fermions, then we 
define 
\begin{eqnarray}
\Label{3.23}
I|A\> = (-1)^{n_A}|A\>.
\end{eqnarray}
Clearly $ I$ is Hermitian,  $ I^2=1$, and $ I$ commutes with all system 
operators.

We now define {\em restricted system operators} $ x$ by
\begin{eqnarray}
\Label{3.24}
x =\left\{ 	\begin{array}{ll}
   I   	\tilde{x}  \qquad  	 &\mbox{if $ \tilde{x}$ is Fermionic}  	\\
   	   	\tilde{x}  \qquad  	&\mbox{if $ \tilde{x}$ is Bosonic}
\end{array}\right.
\end{eqnarray}
Independently of whether $ \tilde{x}$ is Fermionic of Bosonic, $ x$ {\em 
commutes\/} with all bath operators.  These {\em restricted\/} 
system operators, $ x$, are to be contrasted with the {\em full} 
system operators $ \tilde x$.  The $ \tilde x$ are the more physical 
operators, since they are the ones that turn up in the Hamiltonian.  The 
restricted system operators are necessary when we wish to consider operations 
on the reduced density operator $ \rho_{\rm sys} \equiv \tr{B}{\rho}$.

In order to rewrite the Hamiltonian in terms of the restricted 
operators, it is necessary only to rewrite $ H_{\rm Int}$ as
\begin{eqnarray}
\Label{3.25}
H_{\rm Int} = \rmi\hbar\int^\infty_0 dk\,{\k}(k)
\bigl\{d^\dagger(k)Ic - c^\dagger Id(k)\bigr\}
\end{eqnarray}
\subsubsection{Commutation Relations for $ I(t)$ }
Because $ I$ does not commute with $ H_{\rm Int} $, it is time dependent,
and  we re-write is as $ I(t)$.  
Because $ I(t)$ anticommutes with $ d(k,t)$, $d^\dagger(k,t)$, we 
deduce from (\ref{3.15}) and (\ref{3.16}) that
\begin{eqnarray}\Label{3.29}
[I(t),f_{\rm in}(t)]_+ 	&=&	-\sqrt{\gamma}\,c(t) \,\,
= -\sqrt{\gamma}\,I\tilde c(t)
\\ \Label{3.2901} 
[I(t),f^\dagger_{\rm in} (t)]_+	&=& -\sqrt{\gamma}\,c^\dagger(t)
= -\sqrt{\gamma}\,I\tilde c^\dagger(t)
\end{eqnarray}
The commutators of $ I(t)$ with either restricted system operators $ 
x(t)$ or the full system operators $ \tilde{x}(t)$ are zero.  The 
fact that $ I(t)$ does not commute with $ f_{\rm in}(t)$ or 
$ f^\dagger_{\rm in}(t)$, and is time dependent, makes the quantum 
Langevin equations for the restricted operators, (\ref{3.28}), in 
general quite different from those for the full operators, (\ref{3.19}).

However, because $ I(t)^2 = 1$, and because $ I(t)$ commutes with 
system operators of either kind, all equal times algebraic relations 
between different system operators are the same for both full and 
restricted system operators.

\subsubsection{Langevin equation for $ I(t)$}
Using the commutation relations 
(\ref{3.29},\ref{3.2901}),  the equation of motion for $ I(t)$ can be 
deduced by much the same reasoning as above, to be
\begin{eqnarray}
\Label{3.26}
\dot I &=& 
-2\sqrt{\gamma}\left\{f^\dagger_{\rm in}(t) c + c^\dagger  f_{\rm in}(t)
+\sqrt{\gamma}\, I c^\dagger c
\right\} ,
\\ \Label{3.27}
&=& 
-2\sqrt{\gamma}\left\{f^\dagger_{\rm in}(t) I \tilde c + 
\tilde c^\dagger I  f_{\rm in}(t)
+\sqrt{\gamma}\,I\tilde c^\dagger \tilde c
\right\} .
\end{eqnarray}
Finally, notice that $ I(t)^2 = 1$ implies that $ I \dot I + \dot I 
I = 0$.  This can be explicitly shown from (\ref{3.26}) by using 
(\ref{3.29},\ref{3.2901}).

The initial condition, that is the operator  $ I(t_0)$, is determined by the 
formulae (\ref{3.22},\ref{3.23})  using $ d(k,t_0)$, $d^\dagger(k,t_0)$, and by 
definition $ f_{\rm in}(t)$, $ f^\dagger_{\rm in}(t)$, are linear functions of 
$ d(k,t_0)$, $d^\dagger(k,t_0)$.  Thus, we can say that
\begin{eqnarray}\Label{3.2701}
[I(t_0),f_{\rm in}(t)]_+ =[I(t_0),f^\dagger_{\rm in}(t)]_+ =0.
\end{eqnarray}

\subsubsection{Langevin equations for the restricted system operators}
Similar reasoning can be used to deduce that for {\em all\/} 
restricted system operators $ a$, the quantum Langevin equations 
take the form
\begin{eqnarray}\fl \Label{3.28}
\dot{a} = -{\rmi\over \hbar }[a,H_{\rm sys}]   	&-&[a,c^\dagger ]
\left\{{\gamma\over 2 }c + \sqrt{\gamma}\,I f_{\rm in}(t)\right\}
+\left\{{\gamma\over  2} 
c^\dagger +\sqrt {\gamma}\, f^\dagger_{\rm in}(t)I\right\}[a,c].
\end{eqnarray}
This equation can also be deduced by substituting $ \tilde a = a$, or $  Ia$ 
as the case may be, into (\ref{3.19}), and using (\ref{3.27}), together 
with  (\ref{3.29}).

The major advantage of the use of restricted system operators is 
that they are truly independent of the bath operators evaluated at 
the same time.  The whole burden of antisymmetrization required 
between bath and system is borne by the operator $ I(t)$, which is 
{\em not\/} purely a function of $ f_{\rm in}(t)$, 
$f^\dagger_{\rm in}(t)$, but is a dynamical variable whose equation of motion 
must be considered alongside that of the system operators $ a$---namely 
the restricted operator quantum Langevin equation (\ref{3.28}).

\subsubsection{Fermion conservation superselection rule}%
\Label{superselect}
Fermions can only be created and destroyed in pairs.  Hence if 
$ N_B$ is the number of Fermions in the bath, and $ N_{\rm sys}$ is the number 
contained in the system, the quantity
\begin{eqnarray}\Label{3.2801}
K &\equiv & (-1)^{N_B +N_{\rm sys}}
\end{eqnarray}
is a conserved quantity.  Furthermore, the quantity 
\begin{eqnarray}\Label{3.2802}
J(t) &\equiv & (-1)^{N_{\rm sys}}
\end{eqnarray}
is a {\em Bosonic} system variable, which therefore possesses the properties
\begin{eqnarray}\Label{3.2803}
I(t)J(t) = J(t)I(t) &=& K,
\\ \Label{3.280301}
[g(t),J(t)]_\pm &=& 0
\end{eqnarray}
In the latter equation $ g(t)$ is an arbitrary restricted system operator, and 
the choice of anticommutator or commutator  depends on whether the 
corresponding full system operator operator $\tilde g(t)$ is Fermionic ($ +$) 
or Bosonic ($ -$).

This relationship means that correlation functions involving Fermionic full 
system operators can be evaluated using only the restricted system operators 
and 
the conserved operator $ K$.  Thus for example, noting that $\tilde c(t)$ is 
Fermionic,
\begin{eqnarray}\Label{3.2804}
\langle \tilde c^\dagger (t)\tilde c(t')\rangle&=& 
\langle I(t) c^\dagger (t)I(t') c(t')\rangle
\\ \Label{3.2805}
&=& \langle c^\dagger (t)I(t)I(t') c(t')\rangle
\\ \Label{3.2806}
&=& \langle  c^\dagger (t)J(t)K^2J(t') c(t')\rangle
\\ \Label{3.280701}
&=& \langle \breve c^\dagger (t)\breve c(t')\rangle  .
\end{eqnarray}
where we have defined the notation 
\begin{eqnarray}\Label{3.2808}
\breve c &\equiv & J c,
\end{eqnarray}
which amounts in practice to a restricted operator expression of the full 
system operator $ \tilde c$.  Only if there are an odd number of Fermionic 
system operators in the correlation function does the nature of $ K$ come into 
play, and then we find the relation between the $ \tilde c$ correlation 
functions and those involving $ \breve c$ involves a plus or minus sign, 
depending on whether the total number of Fermions in the system is odd or 
even. Such correlation functions are therefore not likely to be of much 
physical relevance, and in fact involve interfering states with different total 
numbers of Fermions; a violation of the superselection rule.
\subsubsection{Properties of solutions}
The solutions of (\ref{3.26},\ref{3.28}) at time $ t$ depend on the initial 
conditions at time $ t_0$, and $ f_{\rm in}(t')$,  $ f_{\rm in}(t')$ for
$ t_0 \le t' \le t$.  In the same way as we derived (\ref{3.2701}), we can show 
that for all $ a(t)$,
\begin{eqnarray}\Label{3.30}
[a(t_0),f_{\rm in}(t)] = [a(t_0),f^\dagger_{\rm in}(t)] =0.
\end{eqnarray}
From the Langevin equations (\ref{3.26},\ref{3.28}) we can show that
\begin{eqnarray}\Label{3.31}
[a(t'),f_{\rm in}(t)] \,\,\,\,= [a(t'),f^\dagger_{\rm in}(t)] =0
&&\quad\mbox{for } t_0 \le t' \le t ;
\\ \Label{3.32}
[I(t'),f_{\rm in}(t)]_+ =[I(t'),f^\dagger_{\rm in}(t)]_+ =0
&&\quad\mbox{for } t_0 \le t' \le t .
\end{eqnarray}
The proof is quite simple; assume the results are true for some 
value $ s$ of $ t'$. Using the Langevin equations evaluate the 
derivative with respect to $ s $ of the left hand sides of
 (\ref{3.31},\ref{3.32}), and then use the relations (\ref{3.31},\ref{3.32})
for $ t'=s$ to show that the results vanish. Hence that the only solution of 
the resultant differential equations corresponds to the truth of  
(\ref{3.31},\ref{3.32}).

In a similar way, one can show that
\begin{eqnarray}\Label{3.33}
[a(t'),f_{\rm out}(t)] \,\,\,\, = [a(t'),f^\dagger_{\rm out}(t)] =0
&&\quad\mbox{for } t \le t' \le t_1 ;
\\ \Label{3.34}
[I(t'),f_{\rm out}(t)]_+ =[I(t'),f^\dagger_{\rm out}(t)]_+ =0
&&\quad\mbox{for } t \le t' \le t_1 .
\end{eqnarray}
Noting now  (\ref{3.2105}), we can derive
\begin{eqnarray}\Label{3.35}
[a(t'),f_{\rm in}(t)] &=& - u(t'-t)\sqrt{\gamma}[a(t'),\tilde c(t)],
\\ \Label{3.36}
[a(t'),f^\dagger_{\rm in}(t)]&=&
- u(t'-t)\sqrt{\gamma}[a(t'),\tilde c^\dagger(t)],
\\ \Label{3.37}
[I(t'),f_{\rm in}(t)]_+ &=& - u(t'-t)\sqrt{\gamma}[I(t'),\tilde c(t)],
\\ \Label{3.38}[I(t'),f^\dagger_{\rm in}(t)]_+ &=&
- u(t'-t)\sqrt{\gamma}[I(t'),\tilde c^\dagger(t)].
\end{eqnarray}
From these it also follows that for any {\em Fermionic} full system 
operator $ \tilde g $
\begin{eqnarray}\Label{3.39}
[\tilde g (t'),f_{\rm in}(t)]_+ 
&=& - u(t'-t)\sqrt{\gamma}[\tilde g (t'),\tilde c(t)]_+,
\\ \Label{3.40}
[\tilde g (t'),f^\dagger_{\rm in}(t)]_+&=&
- u(t'-t)\sqrt{\gamma}[\tilde g (t'),\tilde c^\dagger(t)]_+.
\end{eqnarray}
These results are derived by writing $ \tilde g(t) = g(t) I(t)$, and expanding 
the commutator using (\ref{3.35}--\ref{3.38}).  The corresponding results for a 
{\em Bosonic} full system operator, which is identical with its corresponding 
restricted form, have the same form as (\ref{3.35},\ref{3.36}).

\section {Fermionic Quantum White Noise and Quantum Stochastic 
Differential Equations}
\Label{FQN}
The operators $ f_{\rm in}(t), f^\dagger_{\rm in}(t)$, have the 
idealized anticommutation relations (\ref{3.21}), which leads 
naturally to a formulation of Fermionic quantum white noise.  We 
define a Fermionic Quantum Wiener Process by
\begin{eqnarray}
\Label{4.1}
F(t,t_0) = \int^t_{t_0} f_{\rm in}(t')dt',
\end{eqnarray}
and we assume the averages,
\begin{eqnarray}
\Label{4.2}
\< F^\dagger(t,t_0) F(t,t_0\>  &=& \bar{N}(t-t_0)  	\\
\Label{4.3}
\< F(t,t_0)F^\dagger (t,t_0\>  	&=& (1-\bar{N}) (t-t_0)
\end{eqnarray}
and the anticommutator, from (\ref{3.21})
\begin{eqnarray}
\Label{4.4}
\<[F(t,t_0),F^\dagger(t,t_0)]_+\> = (t-t_0)
\end{eqnarray}
We of course also assume the independence of the $ F$ operators 
defined on non-overlapping time intervals;
\begin{eqnarray}
\Label{4.5}
\< F^\dagger(t,t_0)F(s,s_0)\> &=&\< F(s,s_0) F^\dagger(t,t_0)\>\nonumber\\ 
&=&[F(t,t_0),F(s,s_0)]_+ = 0
\end{eqnarray}
if $ (s,s_0)$ and $ (t,t_0)$ are disjoint.  In frequency space, this 
can be obtained by writing
\begin{eqnarray}
\Label{4.6}
f_{\rm in}(t) = \sqrt{{1\over  2\p}}
\int^\infty_{-\infty} f_0(\w)\rme^{-\rmi\omega t}d\w
\end{eqnarray}
with
\begin{eqnarray}
\Label{4.7}
[f_0(\w),f^\dagger_0(\w')]_+   	&=&\d(\w-\w')  	\\
\Label{4.8}
[f_0^\dagger(\w)f_0(\w')] 	   	&=&\bar{N}\d(\w-\w')   	\\
\Label{4.9}
\< f_0(\w)f_0^\dagger(\w')\>   	&=&(1-\bar{N})\d(\w-\w')
\end{eqnarray}
The $ f_0(\w)$ are therefore like idealized Fermion destruction 
operators, defined on a frequency range $ (-\infty,\infty)$.  In 
practice, the $ d_0(k)$ are the true destruction operators, and the 
correspondence between $ f_0(\w)$ and $ d_0(k)$ is made via the 
relationship (\ref{3.18}), and is only valid for a narrow bandwidth 
around the frequency $ \w_0$.  The components of $ f_{\rm in}(t)$ 
and $ d_{\rm in}(t)$ outside this narrow bandwidth have little 
effect on the solutions of the quantum Langevin equations.

Corresponding to the formula for Bosonic white noise, the density operator 
which gives the averages (\ref{4.2},\ref{4.3})  has the form
\begin{eqnarray}\Label{4.901}
\rho_W(t,t_0) &=& \left(1+\rme^{-\mu}\right)
\exp\left[-{\mu F^\dagger(t,t_0)F(t,t_0)\over t-t_0}\right]
\end{eqnarray}
in which
\begin{eqnarray}\Label{4.902}
\bar N &=& 1\over \rme^\mu +1
\end{eqnarray}

The formulation of $ F(t,t_0)$ as above enables us to develop a 
formal theory of quantum stochastic integration and quantum 
stochastic differential equations which is quite simple and easy to 
use, and whose use gives essentially the same results as any exact 
formulation.  We will also be able to show that the \QSDE s so 
developed are exactly equivalent to the master equation for systems 
interacting with a Fermionic heat bath.
To do this, we partition the time interval $ (t_0,t_f)$, inside which we are 
interested in treating the motion, into subintervals at bounded by times
$ t_0,t_1,t_2,\dots, t_n\equiv t_f$
corresponding to the increments
\begin{eqnarray}\Label{dm3}
\Delta t_i &\equiv & t_{i+1}-t_i
\\ \Label{dm4}
\D F_i &\equiv& F(t_{i+1},t_i) 
 = \int ^{t_{i+1}}_{t_i} f_{\rm in}(t')dt'.
\end{eqnarray}
Depending on whether $ \tilde g(t)$ is a Bosonic or a Fermionic 
operator, $ \D F_i$ will commute or anticommute with $ \tilde g(t_i)$.

The corresponding joint density operator for the Fermi noises partitioned this 
way  is then the direct product
\begin{eqnarray}\fl\Label{dm5}
\rho_{F}&\equiv& \rho_W(t_n,t_{n-1})\otimes
\rho_W(t_{n-1},t_{n-2})\otimes \dots\otimes
\rho_W(t_2,t_{1})\otimes
\rho_W(t_1,t_{0}) .
\end{eqnarray}
This form means that we can consider a trace operation over the Fermion bath
to be taken over each time interval in a discretization; thus if $ Q $ is some 
operator which acts on the fermion bath, we can write
\begin{eqnarray}\fl\Label{dm6}
\tr{B}{Q\rho_{F}}&=&
\tr{(t_n,t_{n-1})}{\tr{(t_{n-1},t_{n-2})}
{\dots \tr{(t_2,t_{1})}{\tr{(t_1,t_{0})}{Q\rho_{F}}}}}.
\end{eqnarray}
This form is particularly useful in deriving correlation function identities.

\subsection{Quantum Stochastic Integration}
As in the case of classical stochastic integration with respect to 
white noise, there are two natural definitions of integration, the 
Ito and Stratonovich methods.  The definition of these is relatively 
straightforward.  
\subsubsection{Fermionic quantum Ito integral}
If  $ \tilde g(t)$ is a  {\em full\/} system operator 
(not a restricted system 
operator), the Fermionic quantum Ito 
integral is defined by
\begin{eqnarray}
\Label{4.10}
\Ito\int^t_{t_0} \tilde g(t')\, dF(t') = \lim_{n\rightarrow\infty}\sum_i \tilde 
g(t_i)
\Delta F_i
\end{eqnarray}
where $ t_0 < t_1 < t_2 < \dots < t_n = t$, and the limit is a 
mean-square limit.  A similar definition is used for 
$ \int^t_t \tilde g(t')\,dF^\dagger(t')$.
The advantage of the Ito definition (\ref{4.10}) of the integral is 
that the increment, $ \Delta F_i$ is seen in the explicit definition on 
the right hand side, to be in the future of $ t_i$. 

In particular, as is the case for classical and Bosonic stochastic integration,
\begin{eqnarray}\Label{4.1002}
\left\langle \Ito\int^t_{t_0} \tilde g(t')\, dF(t')\right\rangle
&=& 0,
\\ \Label{4.1003}
\left\langle \Ito\int^t_{t_0} \tilde g(t')\, dF^\dagger(t')\right\rangle&=& 0.
\end{eqnarray}
Further, depending on whether $ \tilde g(t)$ is Bosonic or Fermionic
\begin{eqnarray}\Label{4.1004}
 \Ito\int^t_{t_0} \big[\tilde g(t'),dF(t')\big]_\pm &=& 0,
\\ \Label{4.1005}
 \Ito\int^t_{t_0} \big[\tilde g(t'),dF^\dagger(t')\big]_\pm &=& 0.
\end{eqnarray}

\subsubsection{Fermionic quantum Stratonovich integral}
The Stratonovich integral can be defined in the same way as it is for Bosonic 
noise \cite{QN}
\begin{eqnarray}\Label{4.1001}
\Strato\int^t_{t_0} \tilde g(t') \,dF(t') = \lim_{n\rightarrow\infty}
\sum_i {\tilde g(t_i)+\tilde g(t_{i+1})\over 2}\,\Delta F_i .
\end{eqnarray}
As in the case of classical and Bosonic noise, we cannot make any connection 
between these two forms of integral without knowing what kind of stochastic 
differential equation is obeyed by the system operators.

Simple relations of the kind (\ref{4.1002}--\ref{4.1005}) do not hold.  
For the first two we need to establish the relationship between the two kinds 
of integral first, and this will be done in Sect.\ref{StratMeans}.

\subsection{Ito quantum stochastic differential equation}
We will define the Ito quantum stochastic differential equation
obeyed by a {\em restricted} system operator $ a$ as
\begin{eqnarray}\Label{4.11}
\Ito da = -{\rmi\over\hbar}[a,H_{\rm sys}]\,dt
&&+{\gamma\over 2}(1-\bar N)
\left\{2c^\dagger a c - ac^\dagger c -c^\dagger  ca\right\}\,dt
\nonumber\\&&
+{\gamma\over 2}\bar N
\left\{2c a c^\dagger - acc^\dagger  -c c^\dagger a\right\}\,dt
\nonumber\\&&
-\sqrt{\gamma}\,[a,c^\dagger]\,I\,dF(t)
+\sqrt{\gamma}\,dF^\dagger(t)\, I\,[a,c].
\end{eqnarray}
This can be written in the alternative form
\begin{eqnarray}
\Label{4.12}
\Ito da =-{\rmi\over\hbar}[a,H_{\rm sys}]\,dt
&+&\left\{{\gamma\over 2}c^\dagger[a,c]-{\gamma\over 2}[a,c^\dagger ]c 
\right\}\,dt
\nonumber\\
&+&{\gamma\over 2}\bar N
\left\{\left[c,[a,c^\dagger]\right]_+ -\left[c^\dagger,[a,c]\right]_+\right\}
\,dt
\nonumber\\&-&
\sqrt{\gamma}\,[a,c^\dagger]\,I\,dF(t)
+\sqrt{\gamma}\,dF^\dagger(t)\, I\,[a,c].
\end{eqnarray}
These definitions are made with foresight; using them we will show how to 
connect 
the Ito and Stratonovich definitions of the \QSDE, ultimately showing that the 
Fermionic Langevin equations as derived above are then to be interpreted as 
Stratonovich 
\QSDE s.

\subsection{Connection between Ito and Stratonovich integrals}
\Label{StratMeans}
We write the  term  $ \tilde g(t_{i+1})$ in
the definition of the Stratonovich integral as
\begin{eqnarray}\Label{4.13}
\tilde g(t_{i+1}) &=& \tilde g(t_i) + \Delta  \tilde g(t_i),
\end{eqnarray}
where $ \Delta  \tilde g(t_i) $ is to be calculated using the Ito \QSDE\ 
(\ref{4.11}).  

Since $ \Delta  \tilde g(t_i) $ is to be used in 
(\ref{4.1001}), only the terms involving the noise operators will be 
significant, since in the limit  terms involving $ \Delta t_i$ as well as a 
noise operator vanish.

The conversion formula will differ, depending on whether
$ \tilde g$ is Fermionic or Bosonic; we will consider first the Fermionic case.
The \QSDE\ (\ref{4.11}) is written for the restricted system operators, but we 
can write 
\begin{eqnarray}\Label{4.14}
\tilde g(t_{i+1}) &=&  I(t_{i+1})g(t_{i+1})
\\ \Label{4.15} &=& K\,J(t_{i+1})g(t_{i+1})
\\ \Label{4.16} &=& K\,\breve g(t_{i+1}).
\end{eqnarray}
Since $ \breve g$ is a restricted system operator, it obeys the \QSDE\
(\ref{4.11}), which means that we can write for the stochastic part of 
$ \Delta  \tilde g(t_i)$
\begin{eqnarray}\fl\Label{4.17}
 \left. \Delta  \tilde g(t_i)\right |_{\rm stochastic}
&=&  K\left\{-\sqrt{\gamma}\, [\breve g,c^\dagger]I\,\Delta F_i
+ \sqrt{\gamma}\,\Delta F^\dagger_iI [\breve g,c]\right\}
\\\fl\Label{4.18}
&=&  K\left\{-\sqrt{\gamma}\, [Jg,c^\dagger]I\,\Delta F_i
+ \sqrt{\gamma}\,\Delta F^\dagger_iI [J g,c]\right\}
\\ \fl\Label{4.19}
&=&  K\left\{-\sqrt{\gamma}\, [g,c^\dagger]_+JI\,\Delta F_i
+ \sqrt{\gamma}\,\Delta F^\dagger_iIJ [ g,c]_+\right\}
\\ \fl\Label{4.20}
&=&  K\left\{-\sqrt{\gamma}\, [g,c^\dagger]_+K\,\Delta F_i
+ \sqrt{\gamma}\,\Delta F^\dagger_iK [ g,c]_+\right\}
\\ \fl\Label{4.21}
&=&  K^2\left\{-\sqrt{\gamma}\, [g,c^\dagger]_+\Delta F_i
- \sqrt{\gamma}\,\Delta F^\dagger_i [ g,c]_+\right\}
\\ \fl\Label{4.22}
&=& -\sqrt{\gamma}\, [g,c^\dagger]_+\Delta F_i
- \sqrt{\gamma}\,\Delta F^\dagger_i [ g,c]_+
\end{eqnarray}
From (\ref{4.2},\ref{4.3}) we can write in stochastic integrals
\begin{eqnarray}\Label{4.23}
\Delta F_i^\dagger \Delta F_i &=& \bar N \Delta t_i,
\\ \Label{4.24}
\Delta F_i \Delta F^\dagger_i &=& (1-\bar N )\Delta t_i,
\end{eqnarray}
and the anticommutation relations also give
\begin{eqnarray}\Label{4.25}
\Delta F_i \Delta F_i &=& \Delta F^\dagger_i \Delta F^\dagger_i =0.
\end{eqnarray}
Carrying out similar reasoning for other integrals, we find that we can write

\subsubsection*{Fermionic integrand $ \tilde g$:}
\begin{eqnarray}\fl\Label{4.26}
\Strato \int _{t_0}^t\tilde g(t') \, dF(t')&=& 
\Ito \int _{t_0}^t\tilde g(t') \, dF(t')-
{\sqrt{\gamma}\,\bar  N\over 2}\int _{t_0}^t[g(t'),c(t')]_+\,dt',
\\\fl \Label{4.28}\Strato \int _{t_0}^t dF(t')\,\tilde g(t') &=& 
\Ito \int _{t_0}^t dF(t')\,\tilde g(t')-
{\sqrt{\gamma}\,(1-\bar  N)\over 2}\int _{t_0}^t[g(t'),c(t')]_+\,dt',
\\\fl\Label{4.27}
\Strato \int _{t_0}^t\tilde g(t') \, dF^\dagger(t')&=& 
\Ito \int _{t_0}^t\tilde g(t') \, dF^\dagger(t')-
{\sqrt{\gamma}\,(1-\bar  N)\over 2}\int _{t_0}^t[g(t'),c^\dagger(t')]_+
\,dt',
\\\fl\Label{4.29}
\Strato \int _{t_0}^t  dF^\dagger(t')\,\tilde g(t')&=& 
\Ito \int _{t_0}^t dF^\dagger(t')\,\tilde g(t')-
{\sqrt{\gamma}\,\bar  N\over 2}\int _{t_0}^t[g(t'),c^\dagger(t')]_+
\,dt'.
\end{eqnarray}
Notice that stochastic integrals use the full operator $ \tilde g$, 
while the correction terms all use the restricted system operators $ 
g,c$.

\subsubsection*{Bosonic integrand $ g$:}In this case the 
restricted operator $ g$ is identical with the full system operator $ 
\tilde g$, so we can write everything in terms of $ g$.
\begin{eqnarray}\fl\Label{4.30b}
\Strato \int _{t_0}^t g(t') \, dF(t')&=& 
\Ito \int _{t_0}^t g(t') \, dF(t')+
{\sqrt{\gamma}\,\bar  N\over 2}\int _{t_0}^t[g(t'),c(t')]\,dt',
\\\fl \Label{4.28b}\Strato \int _{t_0}^t dF(t')\, g(t') &=& 
\Ito \int _{t_0}^t dF(t')\, g(t')-
{\sqrt{\gamma}\,(1-\bar  N)\over 2}\int _{t_0}^t[g(t'),c(t')]\,dt',
\\\fl\Label{4.27b}
\Strato \int _{t_0}^t g(t') \, dF^\dagger(t')&=& 
\Ito \int _{t_0}^t g(t') \, dF^\dagger(t')+
{\sqrt{\gamma}\,(1-\bar  N)\over 2}\int _{t_0}^t[g(t'),c^\dagger(t')]
\,dt',
\\\fl\Label{4.29b}
\Strato \int _{t_0}^t  dF^\dagger(t')\, g(t')&=& 
\Ito \int _{t_0}^t dF^\dagger(t')\, g(t')-
{\sqrt{\gamma}\,\bar  N\over 2}\int _{t_0}^t[g(t'),c^\dagger(t')]
\,dt'.
\end{eqnarray}

\subsection{Stratonovich \QSDE}
We want to write the Stratonovich equivalent of the \QSDE s 
(\ref{4.11},\ref{4.12}), from which we see that  
the Ito integrals to be converted into Stratonovich integrals correspond to
making the choices
\begin{eqnarray}\Label{4.30}
\tilde g(t) \, dF(t) &\to & 
-\sqrt{\gamma}\,[a(t),c^\dagger(t)]\, I(t)\, dF(t),
\\ \Label{4.31}
dF^\dagger(t)\,\tilde  g(t) &\to & \quad\, \sqrt{\gamma}\, dF^\dagger(t)
\,I(t)\,
[a(t),c(t)].
\end{eqnarray}
The combinations $ [a(t),c^\dagger(t)]\, I(t) $ and
$I(t)\,[a(t),c(t)] $ thus form the appropriate Fermionic {\em full} system 
operators.  Using therefore (\ref{4.26},\ref{4.29}), we find the total 
correction term is
\begin{eqnarray}\Label{4.32}
-{\bar N\gamma\over 2}\left\{\big[c,[a,c^\dagger]\big]_+ 
-\big[c^\dagger,[a,c]\big]_+\right\},
\end{eqnarray}
and this is precisely the negative of the third line of (\ref{4.12}), which 
leads to the Stratonovich form 
\begin{eqnarray}\Label{4.33}
\Strato da 
 =-{\rmi\over\hbar}[a,H_{\rm sys}]\,dt
&+&\left\{{\gamma\over 2}c^\dagger[a,c]-{\gamma\over 2}[a,c^\dagger ]c 
\right\}\,dt
\nonumber\\&-&
\sqrt{\gamma}\,[a,c^\dagger]\,I\,dF(t)
+\sqrt{\gamma}\,dF^\dagger(t)\, I\,[a,c].
\end{eqnarray}
This corresponds exactly to the Langevin equation for the restricted system 
operator as given in (\ref{3.28}), and thus justifies the form 
(\ref{4.11},\ref{4.12}) chosen for the Ito \QSDE.

\section{The master equation for the system density operator}
In the Heisenberg picture, the density operator is time independent, and takes 
the form
\begin{eqnarray}\Label{do1}
\rho &=& \rho_{\rm sys}(t_0)\otimes\rho_F.
\end{eqnarray}
This corresponds to an assumption that the initial density operator in the 
Schr\"odinger picture can be factorized into a {\em bath} and a 
{\rm system} term.

Using this density operator, we would say that
\begin{eqnarray}\Label{5.1}
\langle a(t) \rangle &=& \tr{\rm sys}{a\rho_{\rm sys}(t)},
\\ \Label{5.2}
{d\over dt}\langle a(t) \rangle&=& \tr{\rm sys}{a {d\rho_{\rm sys}(t)\over 
dt}}.
\end{eqnarray}
On the other hand, the \QSDE\  (\ref{4.11}) can be used in conjunction with the 
fact that the means of the Ito integrals are zero to show that
\begin{eqnarray}\Label{5.3}
{d\over dt}\langle a(t) \rangle= \bigg\langle
 -{\rmi\over\hbar}[a,H_{\rm sys}]
&&+{\gamma\over 2}(1-\bar N)
\left\{2c^\dagger a c - ac^\dagger c -c^\dagger  ca\right\}
\nonumber\\&&
+{\gamma\over 2}\bar N
\left\{2c a c^\dagger - acc^\dagger  -c c^\dagger a\right\}\bigg\rangle
\end{eqnarray}
from which we deduce, since this equation is true for any restricted system 
operator $ a$
\begin{eqnarray}\Label{5.4}
{d\rho_{\rm sys}\over dt}&=&  {\rmi\over\hbar}[\rho_{\rm sys},H_{\rm sys}]
+{\gamma\over 2}(1-\bar N)\{2 c \rho_{\rm sys}c^\dagger 
- \rho_{\rm sys}c^\dagger c - c^\dagger  c \rho_{\rm sys}\}
\nonumber\\ &&
+{\gamma \bar N\over 2}\{2c^\dagger \rho_{\rm sys} c -
cc^\dagger \rho_{\rm sys} - \rho_{\rm sys}cc^\dagger\}.
\end{eqnarray}

\section{\QQSDE s in the interaction picture}
For a discussion of many aspects of input-output theory it is advantageous to 
formulate an appropriate \QSDE\ theory in the interaction picture, as 
explained 
in Chap.\,11 of \cite{QN} in the case of Bosonic noise.  In such a situation, 
the time-dependent state vectors can be written in terms of the evolution 
operator $ U(t,t_0)$ as 
\begin{eqnarray}\Label{ev0}
|\psi, t\rangle &=& U(t,t_0)|\psi,t_0\rangle .
\end{eqnarray}
In much the same way as for Bosonic noise, we can write a {\em Stratonovich} 
\QSDE\ for the evolution operator corresponding to the Hamiltonian
 (\ref{3.1}) in the form
\begin{eqnarray}\fl\Label{ev6}
\Strato dU(t,t_0) &=& -{\rmi\over\hbar}H_{\rm sys}U(t,t_0) \,dt
+\left\{\sqrt{\gamma}\, dF^\dagger(t)\tilde c +
\sqrt{\gamma}\,dF(t) \tilde c^\dagger\right\}U(t,t_0).
\end{eqnarray}
Our aim now is to transform this to a corresponding Ito form.  We assume there 
exists an Ito equation in the form
\begin{eqnarray}\Label{ev1}
\Ito dU(t,t_0) &=& \left\{ \alpha(t)\, dt + 
\beta(t)\,dF^\dagger (t) + \beta^\dagger(t)\,dF (t)\right\} U(t,t_0),
\end{eqnarray}
and from this derive the relationship between the Stratonovich and Ito 
integrals.

A Stratonovich integral of the evolution operator is defined by
\begin{eqnarray}\fl\Label{ev2}
\Strato \int dF(s)\, U(s) &=&
\lim_{n\to \infty}
\sum \Delta F_i{ U(t_{i+1})+ U(t_i)\over 2}.
\end{eqnarray}
We now express $ U_{i+1}$ in terms of $ U_i$ using the Ito \QSDE\ (\ref{ev1}),
and neglecting terms of order of magnitude $ \Delta t_i^{3/2}$ and higher, we
get
\begin{eqnarray}\fl\Label{ev201}
\Strato \int dF(s)\, U(s)
&=& \lim_{n\to \infty}
\sum \Delta F_i\left\{ 1 + \half\beta_i\Delta F^\dagger_i\right\} U(t_i)
\\ \fl\Label{ev3}
&=& \lim_{n\to\infty}\sum\left\{
 \Delta F_i -\half \beta_i(1-\bar N)\Delta t_i\right\}U(t_i)
\\ \fl\Label{ev4}
&=& \Ito \int dF(s)\, U(s) -\half(1-\bar N) \int \beta(s) U(s)\,ds.
\end{eqnarray}
Similarly
\begin{eqnarray}\fl\Label{ev5}
\Strato \int dF^\dagger(s)\,U(s) &=& 
\Ito \int dF^\dagger(s)\,U(s)  -\half \bar N\int \beta^\dagger(s) U(s)\, ds.
\end{eqnarray}
Using these relations, we can then convert the Stratonovich \QSDE\ to the 
equivalent Ito form 
\begin{eqnarray}\Label{ev7}
\Ito dU(t,t_0) &=&  -\left\{{\rmi\over\hbar}H_{\rm sys}
+{\gamma\bar N\over 2}\tilde c\tilde c^\dagger 
+{\gamma(1-\bar N)\over 2}\tilde c^\dagger\tilde c \right\}U(t,t_0) \,dt
\nonumber \\&&
+\left\{\sqrt{\gamma}\, dF^\dagger(t)\tilde c +
\sqrt{\gamma}\,dF(t) \tilde c^\dagger\right\}U(t,t_0),
\end{eqnarray}
The full density operator at time $ t$ can be written
\begin{eqnarray}\Label{ev8}
\rho(t) &=& U(t,t_0)\rho(t_0) U^\dagger(t,t_0)
\end{eqnarray}
and so obeys the \QSDE\
\begin{eqnarray}\Label{ev9}
d\rho(t) &=& \left\{-{\rmi\over\hbar}[H_{\rm sys},\rho(t)]
-{\gamma\bar N\over 2}[\rho(t),\tilde c\tilde c^\dagger ]_+
-{\gamma(1-\bar N)\over 2}[\rho(t),\tilde c^\dagger\tilde c ]_+\right\}dt
\nonumber\\ &&
-\gamma \tilde c dF^\dagger(t)\rho(t)dF(t)\tilde c^\dagger
-\gamma \tilde c^\dagger dF(t)\rho(t)dF^\dagger(t)\tilde c
\nonumber \\  &&
+\sqrt{\gamma}\left[dF^\dagger(t)\tilde c + dF(t)\tilde c^\dagger,\rho(t)\right]
\end{eqnarray}
\subsection{Alternative derivation of the master equation}
To derive the master equation from the evolution equation 
(\ref{ev9}) still requires us to use the restricted system operators as 
follows.
Firstly, note that the for the product operators 
we have
\begin{eqnarray}\Label{ev10}
\tilde c^\dagger\tilde c &=& c^\dagger c, \qquad
\tilde c\tilde c^\dagger = c c^\dagger ,
\end{eqnarray}
and of course the restricted operator form of $ H_{\rm sys} $ is 
$H_{\rm sys} $ itself.  These operators are therefore proportional to the 
identity in the bath space, and we can write, for example
\begin{eqnarray}\fl\Label{ev11}
\tr{B}{[\rho(t),\tilde c\tilde c^\dagger ]_+}
&=&\tr{B}{[\rho(t), c c^\dagger ]_+}
=[\tr{B}{\rho(t)}, c c^\dagger ]_+
=[{\rho_{\rm sys}(t)}, c c^\dagger ]_+
\end{eqnarray}
However, for the terms on the second line of (\ref{ev9}) this is not 
immediately possible, since $ \tilde c$ and $ \tilde c^\dagger$ do act 
in the bath space, since they do not commute with the noises. We therefore have 
to convert to restricted system operators .  In the interaction picture the 
operator $ I$ introduced in (\ref{3.23}) is time independent, so we can write,
for example
\begin{eqnarray}\Label{ev12}
\tr{B}{\tilde c dF^\dagger(t)\rho(t)dF(t)\tilde c^\dagger}
&=& \tr{B}{ c I\,dF^\dagger(t)\rho(t)dF(t)\,I c^\dagger}
\nonumber \\&=&
 c\tr{B}{  I\,dF^\dagger(t)\rho(t)dF(t)\,I }c^\dagger
\nonumber \\ &=&  
c\tr{B}{ dF(t)\,I^2 \,dF^\dagger(t)\rho(t) }c^\dagger
\nonumber \\ &=&
 (1-\bar N) c\rho_{\rm sys}c^\dagger .
\end{eqnarray}
The trace over the last line of (\ref{ev9}) gives zero, so we derive again the 
master equation in the form (\ref{5.4}).

\subsection{Correlation functions}
From the density operator one can in principle compute all correlation 
functions for the restricted system operators.  Using the relationships
between the operator forms $ \tilde a$, $ \breve a$ and $ a$, as applied in
Sect.\ref{superselect} it is then possible compute the correlation functions 
for the full system operators.

The importance of the full system operators comes from the need to 
compute the correlation functions of the output operators which arises 
because of the relationship (\ref{3.2105}) between outputs, inputs 
and the full system operators.  
\subsubsection{Vacuum input} We can compute
\begin{eqnarray}\Label{5.5}
\langle f^\dagger_{\rm out}(t') f_{\rm out}(t)\rangle
&=& \left
\langle\big( f^\dagger_{\rm in}(t')-\sqrt{\gamma}\,\tilde c^\dagger(t')\big)
\big( f_{\rm in}(t)-\sqrt{\gamma}\,\tilde c(t)\big)\right\rangle .
\end{eqnarray}
The simplest case is if the input field corresponds to the vacuum, in which 
case we get 
\begin{eqnarray}\Label{5.6}
\langle f^\dagger_{\rm out}(t') f_{\rm out}(t)\rangle
&=& 
\gamma\langle\tilde c^\dagger(t')\tilde c(t)\rangle  
\\ \Label{5.601}
&=& 
\gamma\langle\breve c^\dagger(t')\breve c(t)\rangle  
\end{eqnarray}
If we want the number counting  correlation function, then we find in much the 
same way as the Bosonic case that for the {\em time ordered} correlation 
function, in which $ t'>t$
\begin{eqnarray}\Label{5.7}
\langle f^\dagger_{\rm out}(t) f^\dagger_{\rm out}(t') 
f_{\rm out}(t')f_{\rm out}(t')\rangle &=&
\gamma^2
\langle\tilde c^\dagger(t)\tilde c^\dagger(t')\tilde c(t')\tilde c(t)\rangle  
\\ \Label{5.8}
&=& \gamma^2
\langle\breve c^\dagger(t)\breve c^\dagger(t')\breve c(t')\breve c(t)\rangle   
.
\end{eqnarray}

\subsubsection{White noise input}
Let us consider again the correlation function
$ \langle f^\dagger_{\rm out}(t') f_{\rm out}(t)\rangle$ in the case that
the input is non-vacuum, and for $ t<t'$.  In this
case (\ref{5.5}) becomes
\begin{eqnarray}\Label{5.9}
\langle f^\dagger_{\rm out}(t') f_{\rm out}(t)\rangle
&=& \left
\langle\big( f^\dagger_{\rm in}(t')-\sqrt{\gamma}\,\tilde c^\dagger(t')\big)
\big( f_{\rm in}(t)-\sqrt{\gamma}\,\tilde c(t)\right\rangle 
\\ \Label{5.10}
&=&
-\sqrt{\gamma}\left\langle\tilde c^\dagger(t')
\big( f_{\rm in}(t)-\sqrt{\gamma}\,\tilde c(t)\right\rangle 
\\ \Label{5.11}
&=&
-\sqrt{\gamma}\left\langle\tilde c^\dagger(t')
\left\{ dF(t)-\sqrt{\gamma}\,\tilde c(t)\,dt\right\}\right\rangle /dt.
\end{eqnarray}
Here we will consider $ dF(t)$  to be an Ito increment, but there will be no 
difference between an Ito and a Stratonovich version of the increment in this 
formula except when the two times $ t$ and $ t'$ are equal.

Thus it is necessary to compute $ \langle \tilde c^\dagger(t')\,dF(t)\rangle$
where $ t<t'$.  For a general full system operator $ \tilde a(t')$ we
can write  (where $ t_1\equiv t+\Delta t$, for brevity)
\begin{eqnarray}\fl\Label{5.12a}
\langle\tilde a(t')\,\Delta F(t)\rangle=
\nonumber\\ \fl\qquad
\tr{\rm sys}{\tr{(t_1,t')}{\tr{(t,t_1)}{\tilde a
U(t',t_1)U(t_1,t)\Delta F(t)\rho(t)
U^{-1}(t_1,t)U^{-1}(t',t_1)
}}}
\nonumber \\ \fl
\end{eqnarray}
The evolution operator $ U(t',t_1) $ contains no dependence on the noise in the 
interval $(t,t_1) $, so we can than write
\begin{eqnarray}\fl\Label{5.12b}
\langle\tilde a(t')\,\Delta F(t)\rangle=
\nonumber\\ \fl\qquad
\tr{\rm sys}{\tilde a\tr{(t_1,t')}{U(t',t_1)\tr{(t,t_1)}{
U(t_1,t)\Delta F(t)\rho(t)
U^{-1}(t_1,t)
}U^{-1}(t',t_1)}}
\nonumber \\ \fl
\end{eqnarray}
We now write the infinitesimal form 
\begin{eqnarray}\fl\Label{5.12c}
U(t_1,t) \equiv U(t+\Delta t,t)
&=&
-\left\{{\rmi\over\hbar}H_{\rm sys}
+{\gamma\bar N\over 2}\tilde c\tilde c^\dagger 
+{\gamma(1-\bar N)\over 2}\tilde c^\dagger\tilde c \right\}\Delta t
\nonumber \\&&\qquad
+\left\{\sqrt{\gamma}\, \Delta F^\dagger(t)\tilde c +
\sqrt{\gamma}\,\Delta F(t) \tilde c^\dagger\right\}
\end{eqnarray}
Now computing the trace over the interval $ (t,t_1)\equiv(t,t+\Delta t)$, and 
keep only terms of order $ \Delta t$, we get
\begin{eqnarray}\fl\Label{5.13c}
\tr{(t,t_1)}{U(t_1,t)\Delta F(t)\rho(t)U^{-1}(t_1,t)}
&=& \sqrt{\gamma}\bar N[\tilde c,\rho(t)]\Delta t,
\end{eqnarray}
so that 
\begin{eqnarray}\fl\Label{5.13}
\langle\tilde a(t')\,\Delta F(t)\rangle&=&\sqrt{\gamma}\bar N
\tr{\rm sys}{\tilde a\tr{(t_1,t')}{U(t',t_1)
[\tilde c,\rho(t)]
U^{-1}(t',t_1)}}\Delta t
\end{eqnarray}
Taking account of the fact that for $ t'<t$ the average obviously vanishes, 
we get
\begin{eqnarray}\fl\Label{5.14}
\langle \tilde a(t') \,dF(t)\rangle&=& 
\sqrt\gamma\bar N\,u(t'-t) \langle [\tilde a(t'),\tilde c(t)]_\pm\rangle\,dt .
\end{eqnarray}

Using this result we get
\begin{eqnarray}\fl\Label{5.15}
 \langle f^\dagger_{\rm out}(t') f_{\rm out}(t)\rangle&=&
\gamma(1-\bar N) \langle \tilde c^\dagger(t')c(t)\rangle
-\gamma \bar N \langle \tilde c(t) \tilde c^\dagger(t')\rangle
+\bar N\delta(t-t') .
\end{eqnarray}
The result is valid for all $ t,t'$; the case for $ t>t'$ is given by the 
complex conjugate of that for $ t<t'$.

\section{Examples}
\subsection{The two-level ion}
We consider a two level system in which we have a lower energy level with an 
even number of electrons and an upper level with an odd number of 
electrons---hence the terminology ``ion''. The system interacts with an 
electron field, and thus may be described by the choices 
\begin{eqnarray}\Label{6.1}
c&=&\sigma^-,
\\ \Label{6.101}
 c^\dagger &=& \sigma^+,
\\ \Label{6.2}
H_{\rm sys} &=& {\hbar\omega\over 2}\sigma_z,
\end{eqnarray}
so that
\begin{eqnarray}\Label{6.3}
\tilde c&=& I\sigma^- \equiv \tilde \sigma^-,
\\ \Label{6.301}  \tilde c^\dagger &=& I\sigma^+\equiv \tilde \sigma^+,
\\ \Label{6.4}
J &=& \sigma_z,
\\ \Label{6.5}
\breve c&=& J\sigma^-=  -\sigma^-,\qquad  (\breve c)^\dagger = \sigma^+J = -
\sigma^+ .
\end{eqnarray}
\subsubsection*{Quantum Langevin equations for the full system operators}
Using the usual commutation relations in (\ref{3.19}), we get the explicit 
equations of motion
\begin{eqnarray}\Label{6.6}
\dot{\tilde \sigma}^+&=&\quad
 \left(\rmi\omega-{\gamma\over 2}\right){\tilde \sigma}^+ -
\sqrt{\gamma}\,f^\dagger_{\rm in}(t),
\\ \Label{6.7}
\dot{\tilde \sigma}^-&=&
- \left(\rmi\omega+{\gamma\over 2}\right){\tilde \sigma}^- -
\sqrt{\gamma}\,f_{\rm in}(t),
\\ \Label{6.8}
\dot{\tilde\sigma}_z &=& -\gamma(\tilde\sigma_z+1) 
-2\sqrt{\gamma}\,\tilde\sigma^+f_{\rm in}(t) 
-2\sqrt{\gamma}\,\tilde\sigma^-f^\dagger_{\rm in}(t) .
\end{eqnarray}
The equations (\ref{6.6},\ref{6.7}) are linear in $ \tilde \sigma^\pm$, 
$ f_{\rm in}$, $ f^\dagger_{\rm in}$, and are thus exactly solvable.  Although 
the equation for $\tilde \sigma_z$ is not linear, it solution follows from the 
other equations by using identity 
$ [\tilde \sigma^+, \tilde \sigma^-] =\tilde \sigma_z$.

This solvability arises because a two level ion is in fact the same thing as 
another Fermion degree of freedom, that is, in this case the operators 
$ \tilde c$, $ \tilde c^\dagger$ are like Fermion creation and destruction 
operators since $ [\tilde c, \tilde c^\dagger]=1$ and they anticommute with the 
bath Fermion operators.  This is exactly the same situation as for a harmonic 
oscillator interacting with a Bose input field, which is exactly solvable in 
the same way.

\subsubsection*{Stratonovich \QSDE s for the restricted system operators}
These take the form, from (\ref{4.33}) 
\begin{eqnarray}\fl\Label{6.9}
\Strato d\sigma^+ &=&\quad \left(\rmi\omega -{\gamma\over 2}\right)\sigma^+\,dt
+ \sqrt{\gamma}\, dF^\dagger(t)\,(I\sigma_z),
\\\fl \Label{6.10}
\Strato d\sigma^- &=& -\left(\rmi\omega +{\gamma\over 2}\right)\sigma^-\,dt
+ \sqrt{\gamma}\,(I\sigma_z)\, dF(t),
\\\fl \Label{6.11}
\Strato d\sigma_z &=& 
-\gamma(\sigma_z+1)\,dt +2\sqrt{\gamma }\,\sigma^+(I\sigma_z)\,dF(t)
+2\sqrt{\gamma }\,dF^\dagger(t)\,(I\sigma_z)\sigma^- .
\end{eqnarray}
Note that from Sect.\ref{superselect} the quantity which multiplies all the 
noise terms, $ I\sigma_z$ is in this case the conserved operator $ K$.  Thus 
the first two equations are again exactly solvable.

It is surprising to see the apparent difference between the form of the set of
equations (\ref{6.6}--\ref{6.8})  and that of the set 
(\ref{6.9}--\ref{6.11})---however, it must be borne in mind that the 
transformation from the full system operators to the restricted system 
operators 
involves the \emph{time dependent} operator $ I(t)$, and this accounts for the 
apparent difference.  Moreover, once the conserved nature of $ K = I\sigma_z$ 
is 
noted,
one sees that the only difference between the two sets of equations is a sign 
change when the total number of Fermions is odd, which is of course of no 
consequence.

\subsubsection*{Ito \QSDE s for the restricted system operators}
These take the form, from (\ref{4.12}),
\begin{eqnarray}\fl\Label{6.12}
\Ito d\sigma^+ &=& \quad\left(\rmi\omega -{\gamma\over 2}\right)\sigma^+
\,dt
+ \sqrt{\gamma}\, dF^\dagger(t)\,(I\sigma_z),
\\\fl \Label{6.13}
\Ito d\sigma^- &=& -\left(\rmi\omega +{\gamma\over 2}\right)\sigma^-\,dt
+ \sqrt{\gamma}\,(I\sigma_z)\, dF(t),
\\\fl \Label{6.14}
\Ito d\sigma_z &=& 
-\gamma(\sigma_z+1-2\bar N)\,dt +2\sqrt{\gamma }\,\sigma^+(I\sigma_z)\,dF(t)
+2\sqrt{\gamma }\,dF^\dagger(t)\,(I\sigma_z)\sigma^- .
\end{eqnarray}
\subsection{The harmonic oscillator}
The harmonic oscillator coupled to a fermion bath is a problem which cannot be 
solved exactly, in the same way as the two level atom coupled to a Bosonic bath 
is not exactly solvable.
In this case we have harmonic oscillator creation and destruction operators
$ a,a^\dagger$, 
\begin{eqnarray}\Label{6.15}
c&=& a,
\\ \Label{6.16}
 c^\dagger &=& a^\dagger,
\\ \Label{6.17}
H_{\rm sys} &=& {\hbar\omega a^\dagger a},
\end{eqnarray}
so that
\begin{eqnarray}\Label{6.18}
\tilde c&=& I a \equiv \tilde a,
\\ \Label{6.19}  
\tilde c^\dagger &=& I a^\dagger\equiv \tilde a^\dagger,
\\ \Label{6.20}
J &=& (-1)^{a^\dagger a},
\\ \Label{6.21}
\breve a&=& Ja
,\qquad  (\breve a)^\dagger = a^\dagger J .
\end{eqnarray}
\subsubsection*{Quantum Langevin equations for the full system operators}
Using the usual commutation relations in (\ref{3.19}), we get the explicit 
equation of motion
\begin{eqnarray}\fl\Label{6.22}
\dot{\tilde a}&=&
-\rmi \omega \tilde a -\big(2\tilde a^\dagger \tilde a+1\big)
\left\{{\gamma\over 2}\tilde a +\sqrt{\gamma}\, f_{\rm in}(t)\right\}
-\left\{{\gamma\over 2}\tilde a^\dagger +\sqrt{\gamma}\, f^\dagger_{\rm in}(t)
\right\}\big(2\tilde a^2\big).
\end{eqnarray}
\subsubsection*{Stratonovich \QSDE s for the restricted system operators}
We get the form, from (\ref{4.33}) 
\begin{eqnarray}\fl\Label{6.23}
\Strato da &=& -\left(\rmi \omega +{\gamma\over 2}\right)a\, dt 
-\sqrt{\gamma}\,I\,dF(t)
\end{eqnarray}
The difference between the equations for the full and the restricted operators 
is now quite dramatic. The simple form of (\ref{6.23}) is deceptive, since it 
involves the time dependent operator $ I$, whose equation of motion is not 
solvable.

\subsubsection*{Ito \QSDE s for the restricted system operators}
We get, from (\ref{4.12}),
\begin{eqnarray}\fl\Label{6.24}
\Ito da &=& -\left(\rmi \omega +{\gamma(1-2\bar N)\over 2}\right)a\, dt 
-\sqrt{\gamma}\,I\,dF(t)
\\ \fl \Label{6.25}
\Ito d\left(a^\dagger a\right) &=& -\gamma\{(1-2\bar N)a^\dagger a+\bar N\}\,dt
-\sqrt{\gamma}\,a^\dagger I\, dF(t)
-\sqrt{\gamma}\,a \, dF^\dagger(t)\,I
\end{eqnarray}
Using the last equation, we can find the equation of motion for the 
mean number $ n(t)$ to be
\begin{eqnarray}\Label{6.26}
\dot n(t) &=& -\gamma\{(1-2\bar N)n(t)+\bar N\}
\end{eqnarray}
with the stationary solution
\begin{eqnarray}\Label{6.27}
n_{s} &=& {\bar N\over 1- 2 \bar N}.
\end{eqnarray}
For Fermionic noise at a predominant frequency $ \omega$, we know that
\begin{eqnarray}\Label{6.28}
\bar N = {1\over \rme^{\hbar\omega/kT}+1}
\end{eqnarray}
and this yields the correct result for the harmonic oscillator
\begin{eqnarray}\Label{6.29}
n_{s  } &=&{1\over \rme^{\hbar\omega/kT}-1}.
\end{eqnarray}

\section{Conclusions}
This paper answers what is in some sense an academic exercise---how 
do we deal with the Fermion inputs and outputs that are so common in 
the real world, turning up in electronic  and many other systems.  The 
two previous treatments \cite{Sun1999a,Search2002a} gave partial 
answers, the first dealing only with counting statistics, the second 
dealing only with noninteracting particles in cavities.  The 
treatment given here is compatible with both.

As coherent Fermion physics becomes more important, for example in 
highly degenerate trapped cold Fermi vapors \cite{OHara2003a} this 
kind of formalism will undoubtedly become more relevant.  I look 
forward to that time.

\ack I would like to thank Matthew Collett for discussions during the 
beginning of this work, and 
Christian Flindt for helpful comments and questions.
\section*{References}
\bibliographystyle{iopacm}
\bibliography{5Fermion}

\end{document}